\documentclass[12pt]{article}
\usepackage{appendix,graphicx,epstopdf,amssymb,amsmath,mathrsfs}

\newenvironment{keywords}[1][Keywords]{\noindent\textbf{#1:} }{}

\renewcommand\appendix{\par
\setcounter{section}{0}%
\setcounter{subsection}{0}%
\setcounter{table}{0}
\setcounter{figure}{0}
\gdef\thetable{\Alph{table}}
\gdef\thefigure{\Alph{figure}}
\section*{Appendix}
\gdef\thesection{\Alph{section}}
\setcounter{section}{1}}

\begin{document}

\title{Modular Quantum Memories Using Passive Linear Optics and Coherent Feedback}

\author{Hendra~I.~Nurdin\thanks{H. I. Nurdin is with the School of Electrical Engineering and Telecommunications, UNSW Australia, Sydney NSW 2052, Australia. \texttt{Email: h.nurdin@unsw.edu.au}}\, and John~E.~Gough\thanks{J.~E.~Gough is with the Department of Mathematics and Physics, Aberystwyth University, SY23 3BZ, Wales, United Kingdom. \texttt{Email: jug@aber.ac.uk}}}
\date{}

\maketitle \thispagestyle{empty}

\begin{abstract}
In this paper, we show that quantum memory for qudit states encoded in a single photon pulsed optical field has a conceptually simple modular  realization using only passive linear optics and coherent feedback. We exploit the idea  that  two decaying optical cavities can be coupled in a coherent feedback configuration to create an internal mode of the coupled system which is isolated and decoherence-free for the purpose of qubit storage. The qubit memory can then be switched between writing/read-out mode and storage mode simply by varying the routing of certain freely propagating optical fields in the network. It is then shown that the qubit memories can be interconnected with one another to form a qudit quantum memory. We explain each of the phase of writing, storage, and read-out for this modular quantum memory scheme. The results point a way towards modular architectures for complex compound quantum memories. 
\end{abstract}
 

\section{Introduction and Background}
Quantum memories that can store classical and/or quantum information are of fundamental importance in various anticipated quantum information technologies. For example, in quantum communication, quantum memories can be used to store entangled photons and form part of a quantum repeater for long distance distribution of entanglement \cite{qcomm}. In particular, optical quantum memories that operate based on light-matter interaction, have emerged as attractive candidate realizations of quantum memories \cite{LST09}. 

Experimental technologies for optical quantum memories can take on different forms, including  optical delay lines, optical cavities, and trapped atomic ensembles; see \cite{LST09} for a review of  recent developments in optical quantum memories and  \cite{BSARST13} for their prospective applications, such as on-demand single photon sources.  A class of optical quantum memories based on atomic ensembles are  photon-echo memories, which includes the so-called gradient-echo memories (GEM) (also going by the name controlled reversible inhomogeneous broadening (CRIB)).  Modelling of GEM as an infinite cascade of  passive open quantum harmonic oscillators has been reported in \cite{HCHJ13}, using the formalism of quantum stochastic differential equations \cite{HP84,GC85} and the series product for Markovian open  quantum systems \cite{GJ09a}. Theoretical work has characterized the form of the wavepacket required for perfect absorption of a single photon optical state by a two-level system, the so-called rising exponential pulse \cite{perfect-transfer}.  An experiment reported in \cite{excitation-exp} demonstrates  that, in the regime of low average photon numbers,  a  single atom has a higher excitation probability when driven by a coherent state with a rising exponential envelope. Following this, a recent study of linearly coupled networks of open quantum harmonic oscillators (realized as optical cavities or as a collective description of an atomic ensemble) as quantum memories \cite{YJ14} has lead to the formulation of the so-called zero-dynamics principle for perfect storage of a single photon optical state in this class of memories. This principle allowed a simple derivation of the form of the single photon wavepacket for perfect absorption by this class of memories. It turns out that the wavepacket  is also of  rising exponential type, as in \cite{perfect-transfer}.  The present work is partly inspired by the linear quantum memory models studied in \cite{HCHJ13,YJ14}.

Quantum memories as quantum systems are very fragile and are susceptible to unwanted interactions with its ambient environment, resulting in destruction of superposition states prepared in that system, a phenomena known as decoherence. It is widely accepted that some form of control, be it open-loop control techniques such as  dynamical decoupling, optimal control theory, etc, or  quantum feedback control techniques, will be essential to the operation of emerging complex quantum technologies \cite{WM10}; see \cite{BCR10,DP10} for recent surveys of quantum control. 
In the context of quantum memories, in \cite{DLK13} dynamical decoupling has been used to increase the storage time of classical light in an atomic ensemble.  One form of quantum feedback control is coherent feedback \cite{ZJ12}, in which another quantum system, a quantum controller, is attached in a feedback interconnection with the  controlled system without the intervention of any measurements in the feedback loop. To date, various theoretical proposals of coherent feedback for different applications \cite{cfb-theory,JNP08}, as well  as experimental demonstrations  \cite{cfb-exp}, have been reported in the literature.   Coherent feedback will play a key  role  in this work as an enabling tool.
 
In this paper, we propose three principles that lead to a {\em modular} architecture for quantum memories that store qudit states encoded in  pulsed optical fields. The first principle is to create a decoherence-free subsystem  from two  open optical cavity systems, even though individually none of them have a decoherence-free subsystem, by interconnecting them in coherent feedback loop. (After the completion of this work we became aware that Yamamoto had earlier independently arrived at the same idea \cite{Yama14b}. However, our interconnection differs from the Type-1 and Type-2 interconnections studied in \cite{Yama14b}). Storage of light in optical cavities is noted to suffer from limited efficiency due to a trade-off between short cycle of the cavities and long storage, suggesting they may not be suitable as quantum memories \cite{LST09}.  However, the fact that two optical cavities can be connected in feedback to create a decoherence-free subsystem that, in the  ideal situation, is completely protected from decoherence, has not been exploited before. This suggests that the full potential of optical cavities as quantum memories has yet to be explored. The second principle is that the system of two cavities can be readily switched from writing mode and storage mode  for a qubit memory by merely rerouting some propagating fields that interconnect them (with the light being stored in the decoherence free subsystem). Finally, the third principle is that the  basic quantum memory formed using the first and second principles from a module of which several can be connected together to form  a compound quantum memory for  storing qudit states.

The paper exploits the interplay between decoherence-free subsystems, passive linear models of quantum memories, and certain fundamental notions from modern control theory  \cite{Yama14a,YJ14}. It is also partly motivated by a theory of coherent feedback control for disturbance attenuation \cite{JNP08} that was subsequently confirmed  in a tabletop quantum optics experiment in \cite{Mab08}.
Coherent light was injected into one mirror of a bow-tie (optical) cavity, referred to as the {\em plant}, with two partially transmitting mirrors. The cavity plant is coupled in a feedback interconnection to another bow-tie cavity acting as a coherent feedback {\em controller}. When the controller cavity is appropriately designed, broadband destructive interference is engineered resulting in light being blocked out of one of the plant mirrors. This type of control strategy is relevant for quantum memories to minimize leakage of the stored state out of the memory, in accordance with the zero-dynamics principle in \cite{YJ14}. 

\noindent \textbf{Notation.}  $\imath = \sqrt{-1}$, $^*$ denotes the conjugate of a complex number or the adjoint/Hermitian conjugate of a Hilbert space operator. If $X=[X_{jk}]$ is a matrix of complex numbers or  Hilbert space operators then $X^{*}=[X_{kj}^{*}]$, $X^{\#}=[X_{jk}^{*}]$, and $X^{\top}=[X_{kj}]$. ${\rm tr}(\cdot)$ denotes the trace of a matrix or an operator.

\section{Realization of a basic qubit quantum memory}
\label{sec:qubit-mem} 
We begin by illustrating the basic principles for  a single qubit quantum memory using passive linear optics and coherent feedback. This will form the basic modular component that will be extended to a qudit quantum memory in the next section.

The state to be stored is encoded in a ``flying'' continuous-mode single photon qubit of the form $|\psi \rangle =\alpha |\Omega \rangle + \beta |1_{\xi} \rangle$, with $|\alpha|^2+|\beta|^2=1$. Here  $|\Omega \rangle$ denotes the vacuum state of the field (i.e., the vacuum vector on the Fock space of the field) and $|1_{\xi}\rangle$ is a continuous-mode single photon state with temporal wavepacket $\xi$ satisfying $\int_{-\infty}^{\infty}|\xi(s)|^2 ds=1$  \cite{singlephoton}. It is  given by $|1_{\xi} \rangle = B^{*}(\xi) |\Omega \rangle$, where $B^{*}(\xi)=\int_{-\infty}^{\infty} \xi(s) b(s)^*ds$ and $b(t)$ is quantum white noise satisfying the commutation relation $[b(t),b^{*}(s)]=\delta(t-s)$ \cite{GZ04,WaM94,GJ09a,WM10}. Note that $B(\xi)$ satisfies the single mode oscillator commutation relation $[B(\xi),B^{*}(\xi)]=1$ and the temporal wavepacket $\xi$ gives the probability of detecting a single photon at time $t$ as $\int_{-\infty}^t |\xi(s)|^2ds$.

\begin{figure}[tbph]
\centering
\includegraphics[scale=0.60]{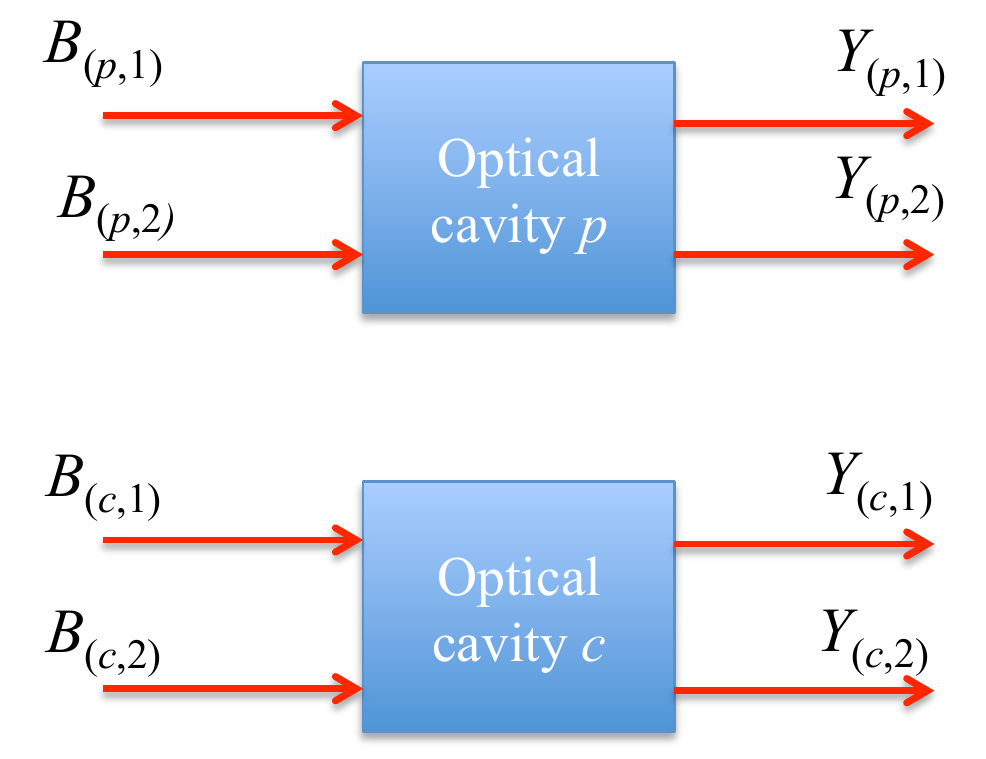}
\caption{(color online) The plant (top) and controller optical cavities with two pairs of input and output ports.}
\label{fig:cavities}
\end{figure}

Consider two optical cavities with two partially transmitting mirrors which, following \cite{JNP08}, we refer to as the ``plant'', denoted with a subscript $p$, and the ``controller'', denoted by the subscript $c$; see Fig.~\ref{fig:cavities}.  Each optical cavity $j \in \{p,c\}$ has two pairs of input and output ports (physically realized as partially transmitting mirrors) through which a propagating optical field can enter and leave the cavity. Optical fields  can  enter through the top port labelled with the subscript $(j,1)$ and exit through the output port  labelled with the same subscript, and through the bottom port labelled with the subscript $(j,2)$ and exit through the  output port  labelled with the same subscript. The field entering through the port  $(j,k)$ has annihilation operator $B_{(j,k)}(t)=\int_0^{t}  b_{{\rm in},(j,k)}(s)ds$, while the field exiting the same port has annihilation field operator $Y_{(j,k)}(t)=\int_0^{t}  b_{{\rm out},(j,k)}(s)ds$, with $b_{{\rm in},(j,k)}(t)$  (resp., $b_{{\rm out},(j,k)}(t)$ ) the associated quantum white noise process for the incoming (resp., outgoing) field into (resp., out of) port $(j,k)$.

With the index $j \in \{c,p\}$, the Heisenberg-Langevin equation in a rotating frame  with respect to the cavities' resonance frequency (taken to be identical for both cavities) for the plant and controller is given by the well-known quantum stochastic differential equation (QSDE)\footnote{We will often omit the time variable $t$ for simplicity.}, see \cite{HP84,GC85,KRP92,GZ04,WaM94}:
\begin{eqnarray*}
da_j &=& -\frac{\gamma_j}{2} a_j dt -\sqrt{\kappa_{j,1}}dB_{(j,1)} -\sqrt{\kappa_{j,2}} dB_{(j,2)},\\
dY_{(j,1)} &=& \sqrt{\kappa_{j,1}} a_j dt + dB_{(j,1)},\\
dY_{(j,2)} &=& \sqrt{\kappa_{j,2}} a_j  dt + dB_{(j,2)},
\end{eqnarray*}
where $\gamma_{j} =\kappa_{j,1}+\kappa_{j,2}$ is the total decay rate for cavity $j$, $a_j$ is the cavity mode for the plant ($j=p$) or controller ($j=c$) satisfying the canonical commutation relations $[a_j,a_k]=0$ and $[a_j,a_k^*]=\delta_{jk}$, with $\delta_{jk}$ denoting the Kronecker delta.

\begin{figure}[tbph]
\centering
\includegraphics[scale=0.5]{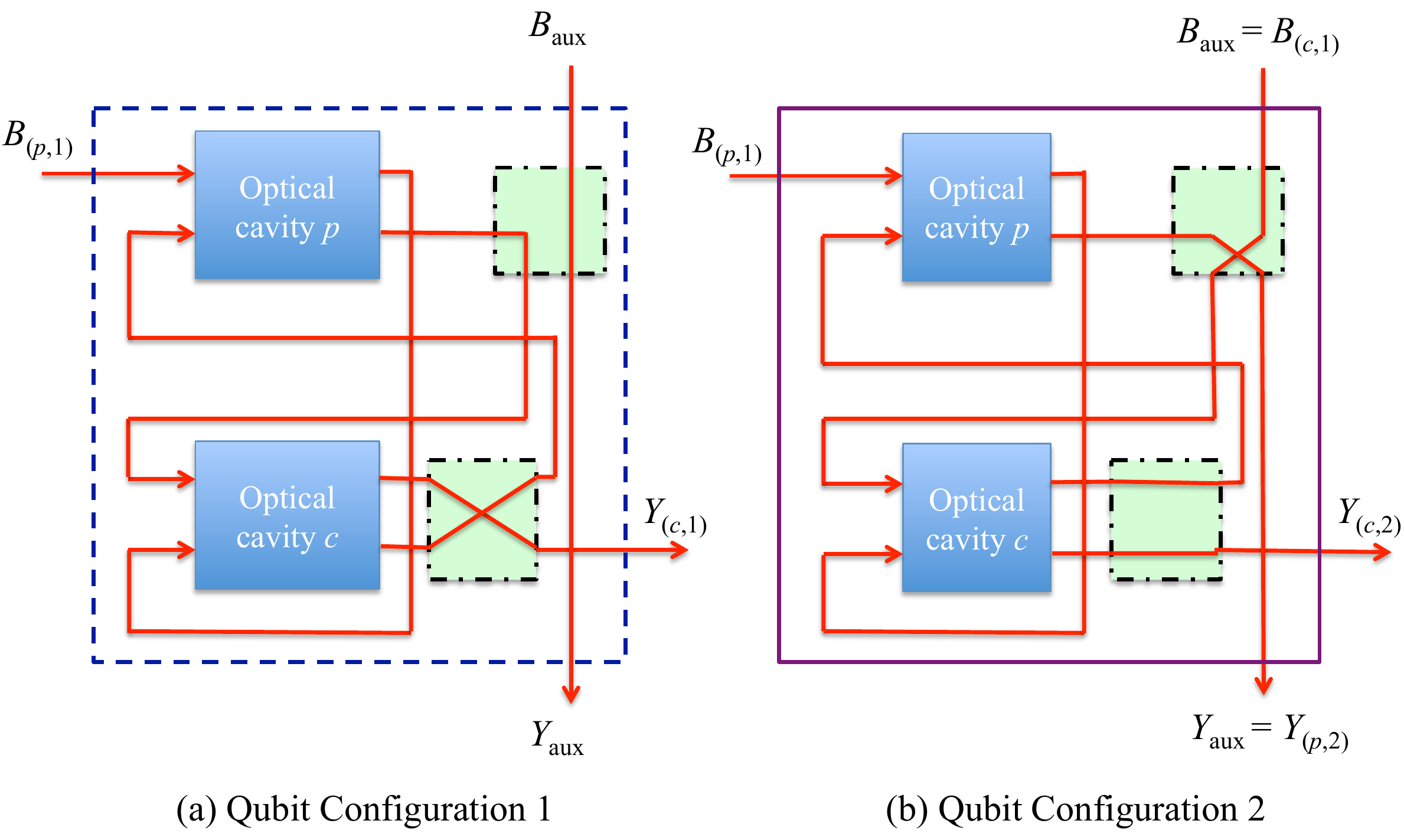}
\caption{(color online) The two configurations of a qubit memory module consisting of the plant and controller optical cavities, (a) Qubit Configuration 1 (left), and (b) Qubit Configuration 2.}
\label{fig:qubitconfigs}
\end{figure}

Set the parameters of the two optical cavities to be identical: $\kappa_{c,j}=\kappa_{p,j}=\sqrt{\gamma/2}$ for $j=1,2$ and hence $\gamma_{c}=\gamma_p=\gamma$. Consider first the interconnection of the plant and controller as shown inside the solid box in  Fig.~\ref{fig:qubitconfigs} (b). We call this Qubit Configuration 2. Note that in the figure we have introduced  an additional   input field $B_{\rm aux}$ and an associated output field $Y_{\rm aux}$ to help make clear and explicit the rerouting of optical fields in the qubit network when the configuration is switched to Qubit Configuration 1 on the left of the figure. The rerouting occurs inside the small dashed dotted green colored boxes in both configurations. The interconnected system then has two outputs $Y_{(p,2)}(t)$ (at the bottom) and $Y_{(c,2)}(t)$ (on the right). 

The Heisenberg-Langevin equation of the closed-loop system in Qubit Configuration 2 is (see Appendix A)
\begin{eqnarray}
\left[ \begin{array}{c} da_p \\ da_c \end{array} \right] 
&=& \left[ \begin{array}{cc} -\gamma/2  & -\gamma/2 \\ -\gamma/2 &  -\gamma/2 \end{array} \right]\left[ \begin{array}{c} a_p \\ a_c \end{array} \right] dt 
-\left[ \begin{array}{cc} \sqrt{\gamma/2}  & \sqrt{\gamma/2} \\ \sqrt{\gamma/2} &  \sqrt{\gamma/2} \end{array} \right]\left[ \begin{array}{c} dB_{(c,1)} \\ dB_{(p,1)} \end{array} \right], \label{eq:QubC1-int}\\
\left[\begin{array}{c} dY_{(p,2)} \\ dY_{(c,2)} \end{array}\right] &=& \left[\begin{array}{cc} \sqrt{\gamma/2}  &  \sqrt{\gamma/2} \\  \sqrt{\gamma/2}  &  \sqrt{\gamma/2} \end{array} \right]\left[\begin{array}{c}  a_c \\ a_p \end{array}\right]dt+\left[ \begin{array}{c}  dB_{(c,1)} \\ dB_{(p,1)} \end{array}\right]. \label{eq:QubC1-out} 
\end{eqnarray}

\begin{figure}[tbph]
\centering
\includegraphics[scale=0.3]{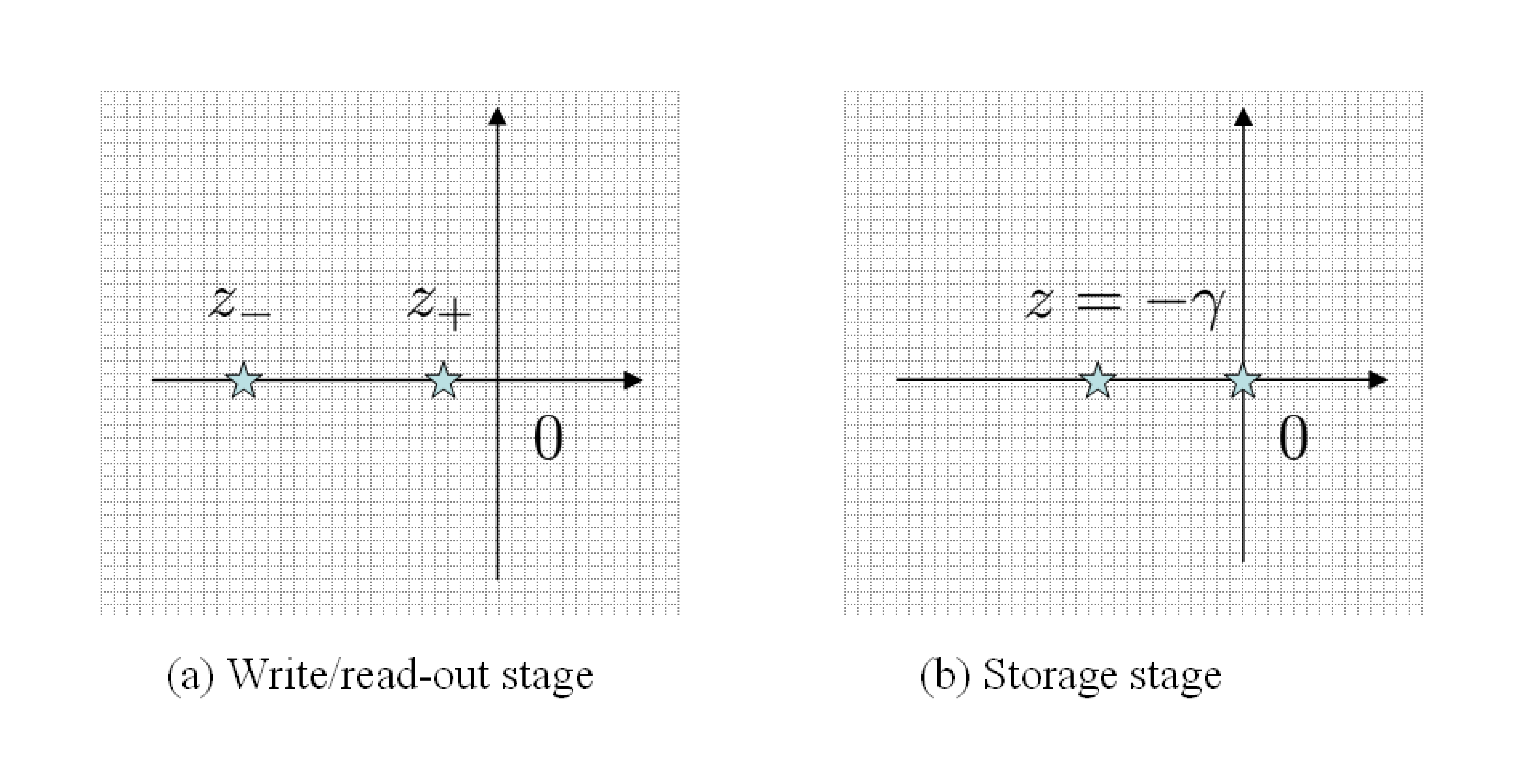}
\caption{(color online) Location of closed-loop eigenvalues : (a) in writing/read-out stage, $z_+=(-1 + \sqrt{3}/2)\gamma$ and   $z_-=(-1 - \sqrt{3}/2)\gamma$, and  (b)  in storage stage, one eigenvalue at $z=-\gamma$ and another at the origin.}
\label{fig:poles}
\end{figure}

Now, observe that the matrix  
$$
A_2= \left[ \begin{array}{cc} -\gamma/2  & -\gamma/2 \\ -\gamma/2 &  -\gamma/2 \end{array} \right],
$$
has eigenvalues  0 and $-\gamma$. That is, $A_2$ has one eigenvalue on the imaginary axis, see Fig.~\ref{fig:poles} (b),  and since the closed-loop system is  a passive linear quantum system it follows from \cite[Lemma 2]{GZ13}, see also \cite[Lemma 3.1 and 3.2]{YG13}, that the system is simultaneously uncontrollable and unobservable (we refer the reader to the Appendix \ref{app:passive_linear} for an overview of passive linear quantum systems and Appendix \ref{app:controllability} for the notions of controllability and observability from modern control theory). It then follows from the result of \cite{Yama14a} that the closed-loop system possesses a decoherence free subsystem (DFS). The physical intuition behind this is clear. Since the system is passive, the cavities will absorb the energy of the incoming optical field and there will either be persistent oscillations with constant energy present, or  the cavities eventually lose all the energy they have absorbed. The former is the case when $A$ has an eigenvalue on the imaginary axis, corresponding to the resonance frequency of some cavity mode, while the latter is when all its eigenvalues are in the left half plane.  In the former case this means energy can be trapped inside the cavity indefinitely, as long as the energy is stored in the appropriate mode. We have showed that coherent feedback can be used to make the oscillator system  have this behaviour, and  identified the associated storage mode. To clarify this, we shall now proceed to explicitly show the DFS. Note that $A_2$ has the decomposition 
$$
A_2 = U  \left[\begin{array}{cc} -\gamma & 0 \\ 0 & 0 \end{array}\right]U^*,
$$
with $U$ the unitary matrix:
$$
U  =\frac{1}{\sqrt{2}}\left[\begin{array}{cc} 1 & 1 \\ 1 & -1\end{array}\right].
$$
Introduce the new rotated modes $\tilde a_p$ and $\tilde a_c$ as 
$$
\left[\begin{array}{c} \tilde a_p \\ \tilde a_c  \end{array}\right]=U^* \left[\begin{array}{c} a_p \\ a_c \end{array}\right].
$$  
That is, $\tilde a_p=\frac{1}{\sqrt{2}}(a_p+a_c)$ and $\tilde a_c=\frac{1}{\sqrt{2}}(a_p-a_c)$. Moreover, since $U$ is unitary, $\tilde a_p$ and $\tilde a_c$ satisfy the same commutation relations as $a_c$ and $a_p$. It follows now that the closed-loop Heisenberg-Langevin equation for the new rotated modes is
\begin{eqnarray}
\left[ \begin{array}{c} d \tilde a_p \\ d \tilde a_c \end{array} \right] &=& \left[ \begin{array}{cc} -\gamma  & 0 \\ 0 &  0 \end{array} \right]\left[ \begin{array}{c} \tilde a_p \\ \tilde a_c \end{array} \right] dt \nonumber   -\left[ \begin{array}{cc} \sqrt{\gamma}  & \sqrt{\gamma} \\ 0 &  0 \end{array} \right]\left[ \begin{array}{c} dB_{(c,1)} \\ dB_{(p,1)} \end{array} \right], \label{eq:QubC1-rint} \\
\left[\begin{array}{c} dY_{(p,2)} \\ dY_{(c,1)} \end{array}\right] &=& \left[\begin{array}{cc} \sqrt{\gamma}  &  0 \\ \sqrt{\gamma} &  0 \end{array} \right]\left[\begin{array}{c}  \tilde a_p \\ \tilde a_c \end{array}\right]dt+\left[ \begin{array}{c}  dB_{(c,1)} \\ dB_{(p,1)} \end{array}\right] \label{eq:QubC1-rout}. 
\end{eqnarray}
and we clearly see  that $\tilde a_c$ is completely decoupled from $\tilde a_p$ and the external optical fields $B_{(c,1)}$ and $B_{(p,1)}$, so this mode represents a decoherence-free subsystem. Moreover, notice that  $\tilde a_c$ also does not appear at all in the outputs $Y_{(p,2)}$ and $Y_{(c,1)}$, a reflection that  the   zero-dynamics principle proposed in \cite{YJ14} is holding. This principle, which is based on the energy balance identity from  \cite{HCHJ13},  states that for perfect storage  in a quantum memory there should be nothing more at the output ports other than vacuum fluctuations.  

We can now describe the three stages of the memory operation. In the time interval $[t_0,t_1]$ ($t_0<t_1$) a quantum state is written into the system via port $(p,1)$. At time $t_1 < t \leq t_2$ the quantum state is stored in the decoherence-free mode $\tilde a_c$ as defined above. Finally, in the read-out stage starting at time $t_2$, the state inside the memory is read-out through the port $(c,1)$.

\subsection{Writing stage ($t_0 \leq t \leq t_1$)} 
\label{sec:qubit-write}

The optical cavities are connected according to Qubit Configuration 1 in the time interval $t_0 \leq t \leq t_1$ and initialized in the ground state $|0_p\rangle |0_c\rangle$ with no photons in either the $p$ or $c$ cavities.  In this configuration, the output $Y_{(p,1)}(t)$ of the plant is passed as the input $B_{(c,2)}(t)$ of the controller, $B_{(c,2)}(t)=Y_{(p,1)}(t)$,   the output $Y_{(c,2)}(t)$ of the controller is passed as the input $B_{(p,2)}(t)$ of the plant, while the output $Y_{(p,2)}$ of the plant is passed as the input $B_{(c,1)}$ of the controller. Notice that this connection is merely  a {\em double pass} scheme through both cavities: first a pass through cavity $p$ then a pass through cavity $c$ followed by a second pass through cavity $p$ and concluded by a final second pass through cavity $c$. A continuous-mode optical field state $|\psi \rangle =\alpha |\Omega \rangle + \beta |1_{\xi} \rangle$ with a certain temporal wavepacket $\xi$ is injected through the port $(p,1)$. The wavepacket $\xi$ is chosen so that this state will get stored in the decoherence-free mode. To see what this wavepacket is, we look at the evolution of the modes $\tilde a_p$ and $\tilde a_c$ as defined previously, but now under Qubit Configuration 1. In this configuration, the Heisenberg-Langevin equation is then (see Appendix A)
\begin{eqnarray*}
\left[ \begin{array}{c} da_p \\ da_c \end{array} \right] &=& \left[ \begin{array}{cc} -\gamma   & -\gamma/2 \\ -3\gamma/2 &  -\gamma  \end{array} \right]\left[ \begin{array}{c} a_p \\ a_c \end{array} \right] dt -\left[ \begin{array}{c} \sqrt{2\gamma} \\ \sqrt{2\gamma}\end{array} \right]dB_{p,1},\\
Y_{c,1} &=& \left[\begin{array}{cc} \sqrt{2\gamma} & \sqrt{2\gamma} \end{array} \right]    \left[ \begin{array}{c} a_p \\ a_c \end{array} \right]  dt+dB_{p,1}.
\end{eqnarray*}
Let $A_1$ be 
$$
A_1 =  \left[\begin{array}{cc} -\gamma  & -\gamma/2 \\ -3\gamma/2 & -\gamma \end{array}\right].
$$
This matrix has two eigenvalues $z_{\pm}=-\gamma \pm \sqrt{3}\gamma/2$ that are all on the left half plane (i.e. all real part of the eigenvalues are negative), so $A_1$ is a Hurwitz matrix, see Fig.~\ref{fig:poles} (a). Again, since the system is passive the system is both controllable and observable. Therefore, this interconnection cannot have a  decoherence free subsystem. From this we can immediately obtain the Heisenberg evolution equation for the modes $\tilde a_p$ and $\tilde a_c$:
\begin{eqnarray*}
\left[ \begin{array}{c} d\tilde a_p \\ d\tilde a_c \end{array} \right] &=& \underbrace{\left[ \begin{array}{cc} -2\gamma   & -\gamma/2 \\  \gamma/2 &  0 \end{array} \right]}_{\tilde A_1}\left[ \begin{array}{c} \tilde a_p \\ \tilde  a_c \end{array} \right] dt \underbrace{-\left[ \begin{array}{c} \sqrt{2\gamma} \\ 0 \end{array} \right]}_{\tilde B_1}dB_{p,1},\\
Y_{c,1} &=& \underbrace{\left[\begin{array}{cc} \sqrt{2\gamma} & 0 \end{array} \right]}_{\tilde C_1}\left[ \begin{array}{c} \tilde a_p \\ \tilde  a_c \end{array} \right]dt+dB_{p,1}.
\end{eqnarray*}

Let $\mathbf{1}(t)$ denotes the Heaviside step function which takes the value 1 for $t \geq 0$ and 0 for $t <0$. Following the results in \cite[Section 5.1]{YJ14}, to write to the decoherence-free mode $\tilde a_c$ the wavepacket  $\xi$ must be of a {\em rising exponential type} (in general a rising exponential with some envelope) given by the second component of the vector $\tilde{\nu}=(\tilde{\nu}_1,\tilde{\nu}_2)^{\top}$,
$$
\tilde{\nu}(t) = -e^{-\tilde A_{1}^{\sharp} (t_1-t)} \tilde C_{1}^{\top}\mathbf{1}(t_1-t).
$$
That is, $\xi(t) = \tilde{\nu}_2(t)$.

Note that for perfect transfer of the single photon into a one photon Fock state in the mode $\tilde a_c$, it is required that $t_0=-\infty$. That is, the rising exponential should begin in the infinite past and stop at $t=t_1$. In  practice, $t_0$ has to be a finite but relatively large negative number (relative to the cavity decay rate), and in this case there will be an unavoidable and small probability that the mode does not perfectly absorb the incoming state if $t_0> -\infty$. 

\subsection{Storage stage ($t_1 < t < t_2$)} 
\label{sec:qubit-store}
At time $t=t_1$ the incoming state $|\psi \rangle$ has been transferred to the mode $\tilde a_c$. Immediately after, the optical fields connecting the cavities are rerouted to make Qubit Configuration 2.  Note that in this configuration, $B_{\rm aux}(t)=B_{(c,1)}(t)$ and $Y_{\rm aux}(t)=Y_{(p,2)}(t)$. The closed-loop equation is as given by (\ref{eq:QubC1-int})-(\ref{eq:QubC1-out}) for the modes $a_p$ and $a_c$, and (\ref{eq:QubC1-rint})-(\ref{eq:QubC1-rout}) for the rotated modes $\tilde a_p$ and $\tilde a_c$. Since the stored state is in the rotated mode $\tilde a_c$ we will now see an expression for the actual state in which the single photon state from the optical field is actually stored.  Now, consider two distinct oscillator modes $a'_1$ and $a'_2$ both initialized in the vacuum state $|0_1\rangle |0_2 \rangle$ having a Heisenberg evolution identical manner to $\tilde a_p$ and $\tilde a_c$, respectively:  
\begin{eqnarray}
\left[ \begin{array}{c} d  a'_1 \\ d a'_2 \end{array} \right] =\left[ \begin{array}{cc} -\gamma  & 0 \\ 0 &  0 \end{array} \right]\left[ \begin{array}{c}  a'_1 \\  a'_2 \end{array} \right] dt \nonumber  \left[ \begin{array}{cc} \sqrt{\gamma}  & \sqrt{\gamma} \\ 0 &  0 \end{array} \right]\left[ \begin{array}{c} dB_{(c,1)} \\ dB_{(p,1)} \end{array} \right].
\end{eqnarray}
If the wavepacket of the incoming field is $\xi(t) =\tilde \nu_2(t)$ as given in Section \ref{sec:qubit-write} then as shown in \cite[Section 5.1]{YJ14}, the modes $a^\prime_1$ and $a^\prime_2$ will converge in the steady-state (as $t \rightarrow \infty$) to the state $|0_1 \rangle |1_2\rangle$. That is, at steady-state mode $a^\prime_1$ contains no photon and $a^\prime_2$ contains one photon. Now, let us return to the modes $\tilde a_p$ and $\tilde a_c$. The modes are initialized in the state $| 0_p \rangle | 0_c \rangle$ (no photons in the two cavities). Then in steady state, identically to the modes $a^\prime_1$ and $a^\prime_2$, the system will be in the state $|\tilde 0_p \tilde 1_c \rangle$ with no photons the mode $\tilde a_p$ and one photon in $\tilde a_c$. This state is given by $|\tilde 0_p \tilde 1_c \rangle=  \tilde a_c^* |0_p \rangle|0_c \rangle=\frac{1}{\sqrt{2}}(a_p^* -a_c^*)|0_p \rangle |0_c\rangle= \frac{1}{\sqrt{2}}|1_p\rangle |0_c\rangle - \frac{1}{\sqrt{2}} |0_p\rangle |1_c\rangle$. Therefore, the one photon is actually stored in an {\rm entangled state} of the two cavities and this is the state that is  preserved in Qubit Configuration 2. Hence, the state $|\psi \rangle=\alpha |\Omega \rangle + \beta |1_{\xi} \rangle$ from the field gets stored as the state $\alpha |0_p \rangle |0_c \rangle +  \frac{\beta}{\sqrt{2}}|1_p\rangle |0_c\rangle - \frac{\beta}{\sqrt{2}} |0_p\rangle |1_c\rangle$.

\subsection{Read-out stage ($t_2 < t$)} 
\label{sec:qubit-read}
To retrieve the stored state after $t=t_2$, the internal optical fields are rerouted again to restore the network to Qubit Configuration 1. The quantum state will leave through the output port $(c,1)$ in the form of the state $|\psi' \rangle=\alpha | 0 \rangle + \beta |1_{\xi'}\rangle$, where the wavepacket $\xi'$ will now be a {\em decaying} exponential type of pulse given by the second component of the vector 
$$
\nu'(t) = (\nu'_1(t),\nu'_2(t))^{\top}=  e^{\tilde A_{1}^{\sharp} (t-t_2)} \tilde C_{1}^{\top}\mathbf{1}(t-t_2).
$$
That is, $\xi'(t) = \nu'_2(t)$, see \cite[Section 5.3]{YJ14}. 

\section{Modular scheme for a qudit quantum memory}
\label{sec:qudit-mem} 

In this section, we show how the qubit quantum memory developed in the previous section forms the basic module for a qudit quantum memory. That is,  we will demonstrate that $n$ single qubit modules can be connected together to form an $n$-level qudit memory to store a single photon optical state of the form $| \psi \rangle =\alpha_0 |\Omega \rangle +  \sum_{k=1}^n \alpha_k |1_{\xi_k} \rangle$ (recall that $|1_{\xi_k} \rangle =B^{*}_k(\xi_k)$). Here, $\alpha_0,\alpha_1,\alpha_2,\ldots,\alpha_n$ are complex numbers satisfying $\sum_{k=0}^n |\alpha_k|^2=1$, and  $\xi_1,\xi_2,\ldots,\xi_n$ are mutually orthogonal normalized temporal wavepackets, $\int_{-\infty}^{\infty} |\xi_j(s)|^2ds=1$ and  $\int_{-\infty}^{\infty} \xi_j(s)^*\xi_k(s) ds = \delta_{jk}$. Note that the qudit state can be used to encode both a quantum state and a classical sequence (i.e., the complex amplitudes $\alpha_0,\alpha_1,\alpha_2,\ldots,\alpha_n$). However, since we can always write  $| \psi \rangle =\alpha_0 |\Omega \rangle + \sqrt{1-\alpha_0^2} |1_{\xi} \rangle$ with $|1_{\xi}\rangle$ a single photon state with wavepacket $\xi =  \sum_{k=1}^n \frac{\alpha_k}{\sqrt{1-\alpha_0^2}} \xi_k $,  it is enough to consider single photon input fields of the form $| \psi \rangle  =  \sum_{k=1}^n \beta_k |1_{\xi_k} \rangle$ with $\sum_{k=1}^n |\beta_k|^2=1$ since the state $|\Omega \rangle$ can be stored in the ground state of the oscillators, while $|1_{\xi} \rangle$  stored in decoherence free modes. Thus we focus on the latter form of $|\psi \rangle$ and come back to the original in Section \ref{sec:qudit-store} 

\begin{figure}[tbph]
\centering
\includegraphics[scale=0.45]{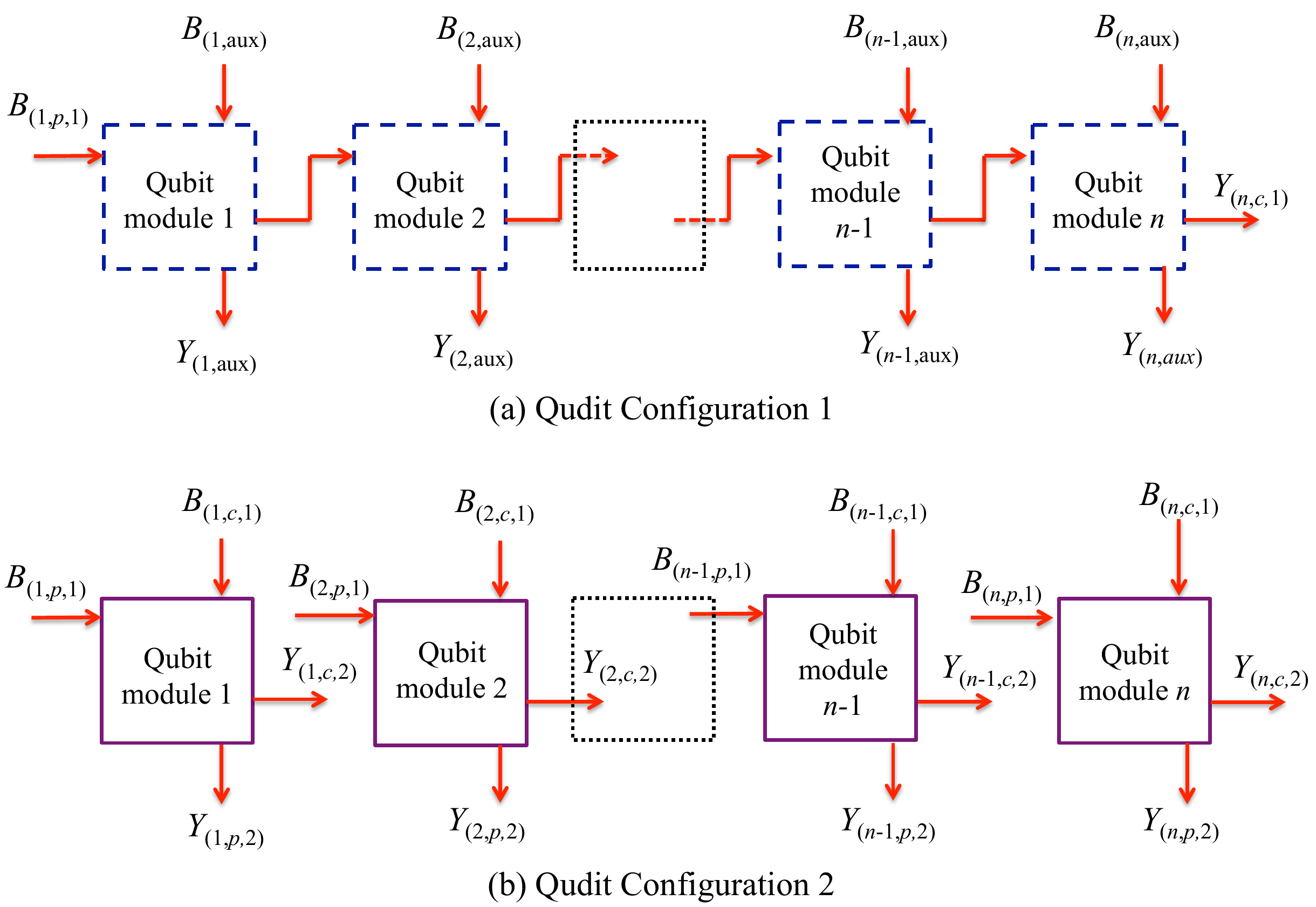}
\caption{(color online) The two configurations of a $n$-level qudit memory  consisting of $n$ qubit memory modules, (a) Qudit Configuration 1 (top), and (b) Qudit Configuration 2 (bottom).}
\label{fig:quditconfigs}
\end{figure}

Each qubit memory module will be labelled as qubit memory $j$, with $j=1,2,\ldots,n$. Thus each qubit module can take on one of the two configurations of Fig.~\ref{fig:qubitconfigs}, corresponding to the dashed and solid boxes. The plant and controller optical cavity modes for module $j$ are denoted as $a_{j,p}$ and $a_{j,c}$, respectively.  The left and top input ports to qubit module $j$ is labelled $(j,p,1)$ and $(j,{\rm aux})$, respectively, while the right and bottom output ports are labelled  $(j,c,k)$ ($k=1,2$  indicating the configuration of the qubit module) and $(j,{\rm aux})$, respectively. However, keep in mind that in Qubit Configuration 2, we have that $B_{(j,{\rm aux})}(t)=B_{(j,c,1)}(t)$ and $Y_{(j,{\rm aux})}(t)= Y_{(j,p,2)}(t)$.

The $n$-level qudit memory consisting of the $n$ qubit memories can be in either of two configurations, which we refer to as  Qudit Configuration 1 and Qudit Configuration 2, see Fig.~\ref{fig:quditconfigs}. In Qudit Configuration 1, Fig.~\ref{fig:quditconfigs} (a), each qubit memory module is in Qubit Configuration 1 (i.e,  none of the qubits possess a decoherence-free subsystem) and they are connected together in a cascade/series connection\footnote{The theory for cascaded systems was initially introduced for optical cavities in \cite{Carm93,Gard93} and  its generalization to general Markovian input-output systems in terms of the series product and quantum feedback network formalisms is given in \cite{GJ09a} and \cite{GJ09b}, respectively. } \cite{Carm93,Gard93,GJ09a,GJ09b} with the output $Y_{(j,c,1)}$ from qubit module $j$ being passed as the input $B_{(j+1,p,1)}$ to qubit module $j+1$, for $j=1,2,\ldots,n-1$. On the other hand, in Qudit Configuration 2, Fig.~\ref{fig:quditconfigs} (b), the $n$ qubit memories are each in Qubit Configuration 2 (each qubit has decoherence-free mode) and are isolated from one  another  (there is no connection between the qubit memory modules). Therefore, the rotated modes $\tilde a_{j,c}$ for each of the qubit modules in Qudit Configuration 2 are individually decoherence-free. 

We now again describe the writing, storage, and read out stage for the $n$-level qudit memory, which operates in an analogous way to the qubit modules.

\subsection{Writing stage ($t_0 \leq t \leq t_1$)} 
\label{sec:qudit-write}
The qudit memory is in Qudit Configuration 1 and all qubit modules are initialized with the $p$ and $c$ modes in their ground state. The single photon pulse $|\psi \rangle$ with  temporal wavepacket  $\xi =\sum_{j=1}^n \alpha_j \xi_j$ is injected through the port for $B_{1,p,1}$ in qubit module 1. The ``sub-wavepackets'' $\xi_1,\xi_2,\ldots,\xi_n$ are chosen so that the associated single photon states $B^{*}(\xi_k)|\Omega \rangle$, $k=1,2,\ldots,n$, will get stored in the decoherence-free modes. To see what they should be, as in the qubit memory case, we look at the evolution of the modes $\tilde a_{j,p}$ and $\tilde a_{j,c}$  in Qudit Configuration 1. In this configuration, the Heisenberg-Langevin equation for $\tilde a_n=(\tilde a_{1,p},\tilde a_{1,c}, \tilde a_{2,p}, \tilde a_{2,c},\ldots, \tilde a_{n,p},\tilde a_{n,c})^{\top}$ is

\begin{eqnarray*}
d\tilde a_n &=& 
\underbrace{\left[ \begin{array}{ccccccccc} -2\gamma   & -\gamma/2 & 0 & 0 & \ldots & 0 & 0 &0 &0 \\ 
 \gamma/2 &  0 & 0 & 0 & \ldots & 0 & 0 & 0 & 0\\
4\gamma & 0 & -2 \gamma & -\gamma/2 &  \ldots & 0 & 0 & 0 & 0\\
0 & 0 & \gamma/2 & 0 & \ldots & 0 & 0 & 0 & 0\\ 
\vdots &  \vdots & \vdots & \vdots & \ddots & \vdots & \vdots & \vdots & \vdots \\
4\gamma & 0 & 4\gamma & 0 & \ldots & -2\gamma & -\gamma/2 & 0 & 0 \\
0 & 0 & 0 & 0 & \ldots & \gamma/2 & 0 & 0 &  0 \\ 
4\gamma & 0 & 4\gamma & 0 & \ldots & 4\gamma & 0 & -2\gamma & -\gamma/2 \\
0 & 0 & 0 & 0 & \ldots & 0& 0 &  \gamma/2 & 0  \end{array} \right]}_{\tilde A_{n,1}} \tilde a_n dt \\
&& \underbrace{-\left[ \begin{array}{ccccccccc} \sqrt{2\gamma} & 0 & \sqrt{2\gamma} & 0 & \ldots & \sqrt{2\gamma} & 0 & \sqrt{2\gamma} & 0 \end{array} \right]^{\top}}_{\tilde B_{n,1}}dB_{p,1},\\
dY_{c,1}&=&\underbrace{\left[\begin{array}{ccccccccc} \sqrt{2 \gamma} & 0 & \sqrt{2 \gamma} & 0 & \ldots & \sqrt{2 \gamma} & 0 & \sqrt{2 \gamma} & 0 \end{array} \right]}_{\tilde C_{n,1}}\tilde a_n dt +dB_{p,1}
\end{eqnarray*}

Since $\tilde A_{n,1}$ is lower $2 \times 2$ block triangular while all of the $2\times 2$ block matrices on its diagonal blocks are Hurwitz, it follows that $\tilde A_{n,1}$ is also Hurwitz.  Hence in this configuration decoherence free modes do not exist. Now, as in the qubit case, to write the incoming input quantum state into the modes $\tilde a_{1,c},\tilde a_{2,c},\ldots, \tilde a_{n,c}$, it again follows from the results of \cite[Section 5.1]{YJ14} that the sub-wavepackets $\xi_1,\xi_2,\ldots,\xi_n$ must be set to even indexed  components $\tilde{\nu}_2,\tilde{\nu}_4,\ldots,\tilde{\nu}_{2n}$ of the vector  $\tilde{\nu}=(\tilde{\nu}_1,\tilde{\nu}_2,\tilde{\nu}_3,\tilde{\nu}_4,\ldots,\tilde{\nu}_{2n-1},\tilde{\nu}_{2n})^{\top}$ given by 
$$
\tilde{\nu}(t)=-e^{-\tilde A_{n,1}^{\sharp} (t_1-t)} \tilde C_{n,1}^{\top}\mathbf{1}(t_1-t).
$$
That is, $\xi_k(t)= \tilde \nu_{2k}(t)$ for $k=1,2,\ldots,n$.

\subsection{Storage stage ($t_1 < t < t_2$)} 
\label{sec:qudit-store}
At time $t=t_1$ the single photon state $|\psi \rangle = \sum_{k=1}^n \beta_k |1_{\xi_k}\rangle$ has been written to the modes $\tilde a_{1,c},\tilde a_{2,c},\ldots, \tilde a_{n,c}$, and the optical fields are immediately rerouted so that the qudit memory is in Qudit Configuration 2 (with each qubit module in Qubit Configuration 1). Each of the single qubit modules are now  decoherence-free  with a single photon stored in the decoherence free mode $\tilde a_c$. Let $|0_{k,j}\rangle$ and  $|1_{k,j}\rangle$ denote the ground and one photon state of the mode $a_{k,j}$. 
Following the discussion in Section \ref{sec:qubit-store}, the optical state is preserved within the oscillators in the decoherence-free entangled state $\sum_{j=1}^n \beta_j \otimes_{k=1}^{n} | \phi_{jk} \rangle$, with  
$$
|\phi_{jk} \rangle = \left\{\begin{array}{cc} |0_{k,p} \rangle |0_{k,c} \rangle, & \hbox{if $k \neq j$},\\
 \frac{1}{\sqrt{2}}|1_{j,p} \rangle |0_{j,c} \rangle - \frac{1}{\sqrt{2}}|0_{j,p} \rangle |1_{j,c} \rangle, & \hbox{if $k=j$}\end{array} \right. 
 $$  
Coming back to optical states of the form $|\psi \rangle = \alpha_{0} |\Omega \rangle + \sum_{j=1}^n \alpha_j |\xi_j \rangle$, the Fock vacuum $|\Omega\rangle $ will trivially be mapped  to the ground state $\otimes_{j=1}^n |0_{j,p}\rangle |0_{j,c}\rangle$, so that $|\psi \rangle$ gets stored in the entangled state $\alpha_0 \otimes_{j=1}^n |0_{j,p}\rangle |0_{j,c}\rangle + \sum_{j=1}^n \alpha_j \otimes_{k=1}^{n} | \phi_{jk} \rangle$.
  
\subsection{Read-out stage ($t_2 < t$)} 
To retrieve the stored state after $t=t_2$, the internal optical fields are rerouted again to restore it to Qudit Configuration 1. The quantum state will leave through the output port of $Y_{(n,c,2)}$ of qubit module $n$, which is the rightmost module in the cascade. Analogous to the discussion in Section \ref{sec:qubit-read}, using the results from \cite[Section 5.3]{YJ14}, the  output wavepacket will be of the form  $|\psi' \rangle=\alpha_{0} | \Omega \rangle + \sum_{k=1}^n \alpha_k B^*(\xi'_k)| \Omega \rangle $, where the sub-wavepacket $\xi'_k$ is the $2k$-th component of the vector $\nu'$ given by
\begin{eqnarray*}
\nu'(t) &=& (\nu'_1(t),\nu'_2(t),\ldots,\nu'_{2n}(t))^{\top} \\
&=&  e^{\tilde A_{n,1}^{\sharp} (t-t_2)} \tilde C_{n,1}^{\top}\mathbf{1}(t-t_2).
\end{eqnarray*}
That is, $\xi'_k(t) = \nu'_{2k}(t)$ for $k=1,2,\ldots,n$. 

\section{Discussion}
Although we  focused on   storage of a propagating single photon wavepacket, the scheme here can also store propagating optical coherent states of the form $|f \rangle = e^{B^{*}(f)-B(f)}=e^{\int_{-\infty}^{\infty} (f(s)b(s)^{*} - f(s)^* b(s))ds}$, in an analogous manner. Here, $f=\sum_{k=1}^n \alpha_k f_k$, where the $f_k$'s are square-integrable, $\|f_k\|^2=\int_{-\infty}^{\infty} |f_k(s)|^2ds<\infty$ for $k=1,2,\ldots,n$, and mutually orthogonal, $\int_{-\infty}^{\infty} f_j(s)^*f_k(s) d =\delta_{jk}$, so that also $f$ is square-integrable and $\int_{-\infty}^{\infty} |f(s)|^2 ds=\sum_{k=1}^n \|f_k\|^2$. The details for coherent state write/storage/read-out we  leave to the reader as it follows mutatis mutandis from the single photon case using the results of \cite{YJ14}. However, note that unlike single photon states, optical coherent states only contain classical information encoded in the coefficients $\alpha_2,\alpha_1,\ldots,\alpha_n$.

Our proposed scheme is dependent on the controller and optical cavities being identical. Deviation of the parameters of each cavity from this ideal situation inevitably leads to some leakage of the stored quantum state from the memory. In any case, practical quantum memories cannot be expected to hold a quantum state indefinitely, but only long enough to complete an information processing task. However, there are ways to mitigate this in order to slow down the leakage. One possible approach is to  insert a disturbance attenuating controller in the style of \cite{JNP08} between the free output port $(c,2)$ and input port $({\rm aux})$ in Qubit Configuration 2 (Fig.~\ref{fig:qubitconfigs} (b)). The effect of imperfections in the implementation to  performance of the memory scheme, and possible methods to reduce these effects,  deserve further investigation but is beyond the scope of the present paper which aims only to give an exposition of the key ideas. One could also consider questions such as, what would be an optimal architecture  of the quantum memory with respect to its robustness to various imperfections in its realization? These are further open topics to be studied in the future.

\section{Conclusion}
\label{sec:conclu}
In this paper, we have shown that one can build a modular quantum memory structure for quantum and/or classical information encoded in single photon or coherent field states using only linear optics and coherent feedback. Three key ideas underlie our proposal: (i) creation of a decoherence-free subspace by coherent feedback,  (ii) distinct configurations for passive optical cavities can be set up for writing/read-out and storage, and switching among these configurations can be achieved by merely rerouting the optical fields that interconnect the cavities, and (iii) qubit memories can be interconnected to form qudit memories.  A decoherence-free configuration is used for storage while a non-decoherence-free configuration allows ``opening up'' the cavities for writing-in the quantum state of the optical field and for retrieving a stored state. We give a concrete architecture of a qubit quantum memory that can store a qubit encoded in an optical field state and show they can be interconnected to form a qudit memory unit. Thus, our approach presents  a way to construct complex quantum memories by interconnecting simpler quantum memory components.

\section*{Acknowledgements}
\noindent
HN acknowledges the support of the Australian Research Council  through the grant DP130104191 and the Visiting Researchers scheme of the Department of Mathematics and Physics, Aberystwyth University, for a 
 visit in June/July 2014.
JG would like to thank the Isaac Newton Institute for Mathematical Sciences, 
Cambridge, for support and hospitality during the programme \textit{Quantum Control Engineering}, July/August 2014,
where work on this paper was undertaken.

\section*{Appendices}
\appendices

\section{The Network Calculations}
\label{app:Configurations}
For Markovian open quantum systems driven by $n$ vacuum noise inputs, the model is
specified by a triple 
\begin{equation*}
\mathbf{G}\sim (S,L,H)
\end{equation*}
referred to as the set of Hudson-Parthasarathy coefficients, or more
prosaically as the ``SLH''. Their roles are to describe the input-to-output
scattering $S= [S_{jk}]$ of the external noise fields $b_{k}\left( t\right) $%
, the coupling $L= [L_j ]$ of the noise to the system, and the internal
Hamiltonian $H$ of the system respectively.

The SLH formalism for quantum Markov models deals with the category of models

\begin{equation*}
S=\left[ 
\begin{array}{ccc}
S_{11} & \cdots & S_{1n} \\ 
\vdots & \ddots & \vdots \\ 
S_{n1} & \cdots & S_{nn}
\end{array}
\right] ,L=\left[ 
\begin{array}{c}
L_{1} \\ 
\vdots \\ 
L_{n}
\end{array}
\right] ,H .
\end{equation*}

These may be assimilated into the \textit{model matrix} 
\begin{eqnarray*}
\mathsf{V}  &=& \left[ 
\begin{array}{cc}
-\frac{1}{2}L^{\ast }L-iH & -L^{\ast }S \\ 
L & S
\end{array}
\right]  \\
&=&  \left[ 
\begin{array}{cccc}
-\frac{1}{2}\sum_{j}L_{j}^{\ast }L_{j}-iH & -\sum_{j}L_{j}^{\ast }S_{j1} & 
\cdots & -\sum_{j}L_{j}^{\ast }S_{jm} \\ 
L_{1} & S_{11} & \cdots & S_{1n} \\ 
\vdots & \vdots & \ddots & \vdots \\ 
L_{n} & S_{n1} & \cdots & S_{nn}
\end{array}
\right]\\
&=& 
\left[ 
\begin{array}{cccc}
\mathsf{V}_{00} & \mathsf{V}_{01} & \cdots & \mathsf{V}_{0m} \\ 
\mathsf{V}_{10} & \mathsf{V}_{11} & \cdots & \mathsf{V}_{1n} \\ 
\vdots & \vdots & \ddots & \vdots \\ 
\mathsf{V}_{n0} & \mathsf{V}_{n1} & \cdots & \mathsf{V}_{nn}
\end{array}
\right] .
\end{eqnarray*}

The system with Hilbert space $\mathfrak{h}$ driven by $n$ independent Bose
quantum processes with Fock space $\mathfrak{F}$ will have a unitary
evolution $V(t)$\ on the space $\mathfrak{h}\otimes \mathfrak{F}$ where $V(t)
$ is the solution to the quantum stochastic differential equation \cite{HP84}

\begin{eqnarray*}
dV(t) &=& \{ (S_{jk}-\delta _{jk})\otimes d\Lambda _{jk}(t)+L_{j}\otimes
dB_{j}^{\ast }(t) \\
&&
  -L_{j}^{\ast }S_{jk}\otimes dB_{k}(t)-(\frac{1}{2}L_{k}^{\ast
}L_{k}+iH)\otimes dt  \} \,V(t)
\end{eqnarray*}

with initial condition $V(0)=I$. (We adopt the convention that repeated
Latin indices imply a summation over the range $1,\cdots ,n$.) Formally, the
Bose noise can be thought of as arising from quantum white noise processes $%
b_{k}(t)$ satisfying a set singular of commutation relations 
\begin{equation*}
\left[ b_{j}(t),b_{k}(s)^{\ast }\right] =\delta _{jk}\delta (t-s),
\end{equation*}
with 
\begin{eqnarray*}
B_{j}(t) &=&\int_{0}^{t}b_{j}(s)ds,\quad B_{j}^{\ast
}(t)=\int_{0}^{t}b_{j}(s)^{\ast }ds, \\
\Lambda _{jk}(t) &=&\int_{0}^{t}b_{j}(s)^{\ast }b_{k}(s)ds.
\end{eqnarray*}
Conditions guaranteeing unitarity are $S=\left[ S_{jk}\right] $ is
unitary, $L=\left[ L_{j}\right] $ is bounded, and $H$ bounded self-adjoint.

\bigskip

\noindent{\textbf{The Network Rules}}

\bigskip

The rules for construction arbitrary network architectures were derived in 
\cite{GJ09b}.

\bigskip

\centerline{\textbf{\# 1 The Parallel Sum Rule}}

\bigskip

\begin{figure}[h]
\centering
\includegraphics[width=0.30\textwidth]{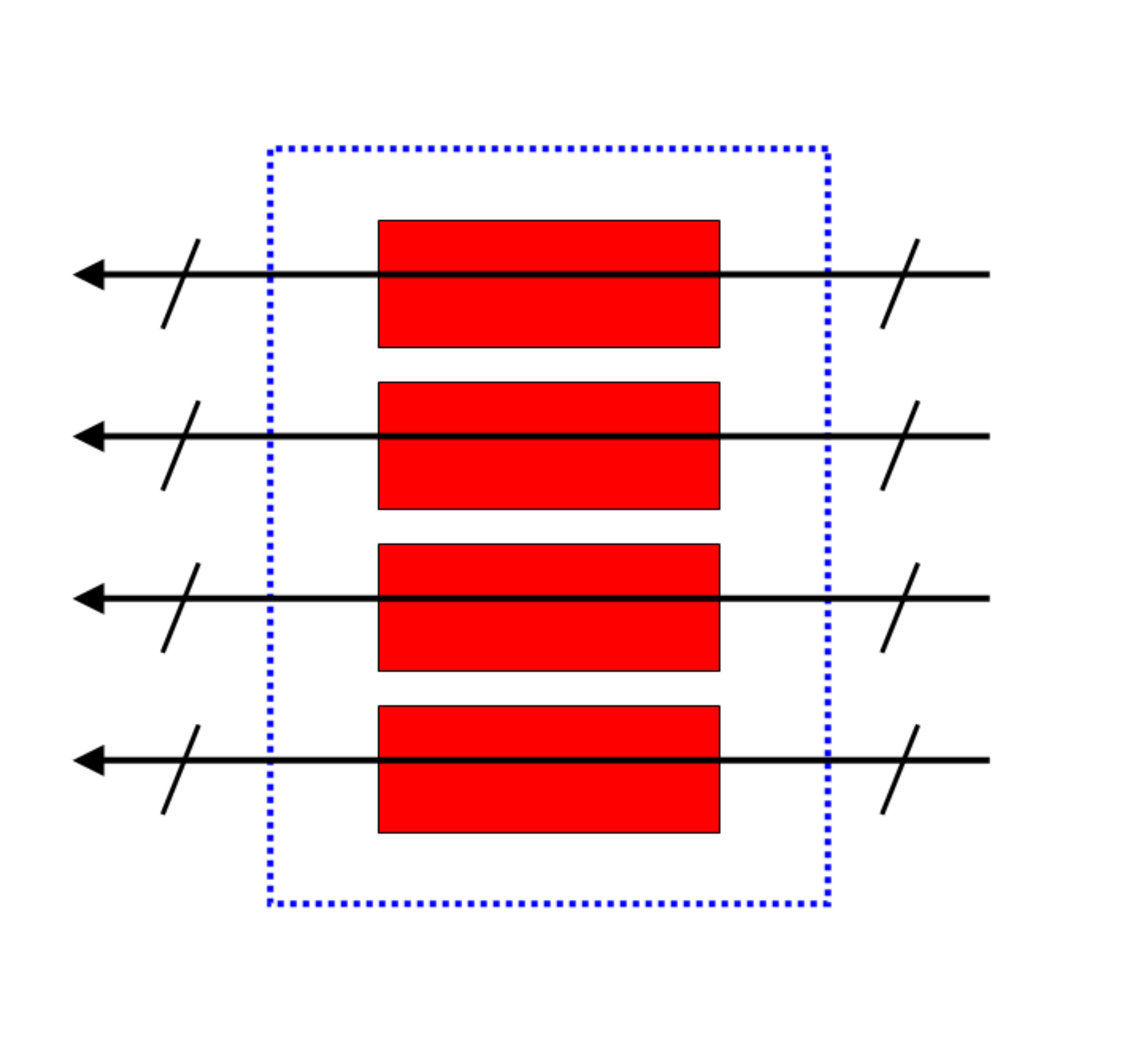}
\caption{(color online) Several SLH models run in parallel: they correspond
to one single SLH model.}
\label{fig:QFN_Parallel}
\end{figure}

If we have several quantum Markov models with independent inputs then they
may be assembled into a single SLH model, see Fig. \ref{fig:QFN_Parallel}. 

\begin{eqnarray*}
 \boxplus _{j=1}^{n} \left( S_{j},L_{j},H_{j}\right) =  \left( \left[ 
\begin{array}{ccc}
S_{1} & 0 & 0 \\ 
0 & \ddots & 0 \\ 
0 & 0 & S_{n}
\end{array}
\right] ,\left[ 
\begin{array}{c}
L_{1} \\ 
\vdots \\ 
L_{n}
\end{array}
\right] ,H_{1}+\cdots +H_{n}\right) .
\end{eqnarray*}

Note that the components need not be distinct - that is, observables
associated with one component are not assumed to commute with those of
others. In this case the definition is not quite so trivial as it may first
appear.

\bigskip

\centerline{\textbf{\# 2 The Feedback Reduction Rule}}

\bigskip

\begin{figure}[h]
	\centering
  	\includegraphics[width=0.40\textwidth]{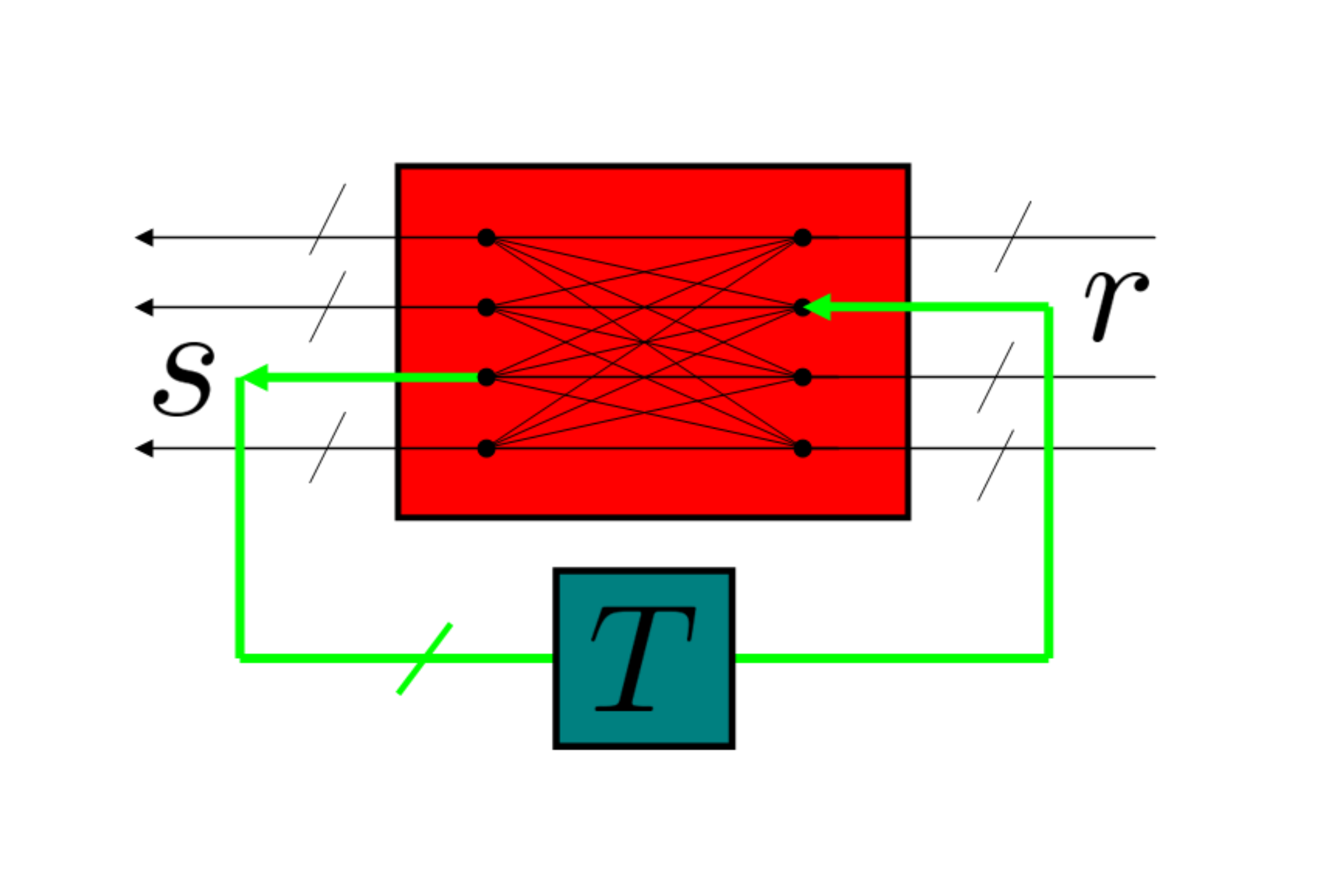}
  	\caption{(color online) We feed selected outputs back in as inputs to get a reduced model. Here the unitary gain $T$ is taken to be the adjacency matrix
		$\eta $	describing the port connections.}
	\label{fig:QFN_FR}
\end{figure}

If we wish to feed an output back in as an input, we obtain a reduced model as depicted in Fig. \ref{fig:QFN_FR}.

We write the operators $\left( S,L,H\right) $ of the network as 
\begin{equation*}
S=\left[ 
\begin{array}{cc}
S_{\mathtt{ii}} & S_{\mathtt{ie}} \\ 
S_{\mathtt{ei}} & S_{\mathtt{ee}}
\end{array}
\right] ,\quad L=\left[ 
\begin{array}{c}
L_{\mathtt{i}} \\ 
L_{\mathtt{e}}
\end{array}
\right] .
\end{equation*}
The feedback reduced model matrix may be conveniently expressed as 
\begin{equation*}
\mathcal{F}\left( \mathbf{V,\eta }^{-1}\right) _{\alpha \beta }\triangleq 
\mathsf{V}_{\alpha \beta }+\mathsf{V}_{\alpha \mathtt{i}}\left( \eta -%
\mathsf{V}_{\mathtt{ii}}\right) ^{-1}\mathsf{V}_{\mathtt{i}\beta }
\end{equation*}
for $\alpha ,\beta \in \left\{ 0,\mathtt{e}\right\} $, where $\eta $ is the
(unitary) \textit{adjacency matrix} 
\begin{equation*}
\eta _{sr}=\left\{ 
\begin{array}{cc}
1, & \text{if }\left( s,r\right) \text{ is an internal channel,} \\ 
0, & \text{otherwise.}
\end{array}
\right.
\end{equation*}
Here the ``gain'' $\eta $ is the set of instructions as to which internal
output port gets connected up to which internal input port. We, of course,
have $\eta =1$ if we match up the labels of the input and output ports
according to the connections, however, it is computationally easier to work
with a general labeling and just specify the adjacency matrix. The reduced
model matrix $\mathbf{V}^{\text{red}}$ obtained by eliminating all the
internal channels is determined by the operators $\left( S^{\text{red}},L^{%
\text{red}},H^{\text{red}}\right) $\ given by 
\begin{eqnarray*}
S^{\text{red}} &=&S_{\mathtt{ee}}+S_{\mathtt{ei}}\left( \eta -S_{\mathtt{ii}%
}\right) ^{-1}S_{\mathtt{ie}}, \\
L^{\text{red}} &=&L_{\mathtt{e}}+S_{\mathtt{ei}}\left( \eta -S_{\mathtt{ii}%
}\right) ^{-1}L_{\mathtt{i}}, \\
H^{\text{red}} &=&H+\sum_{i=\mathtt{i},\mathtt{e}}\mathrm{Im} \left\{L_{j}^{*}S_{j%
\mathtt{i}}\left( \eta -S_{\mathtt{ii}}\right) ^{-1}L_{\mathtt{i}} \right\} .
\end{eqnarray*}

\centerline{\textbf{\# 3 Construction}}

\bigskip

\begin{figure}[htbp]
\centering
\includegraphics[width=0.40\textwidth]{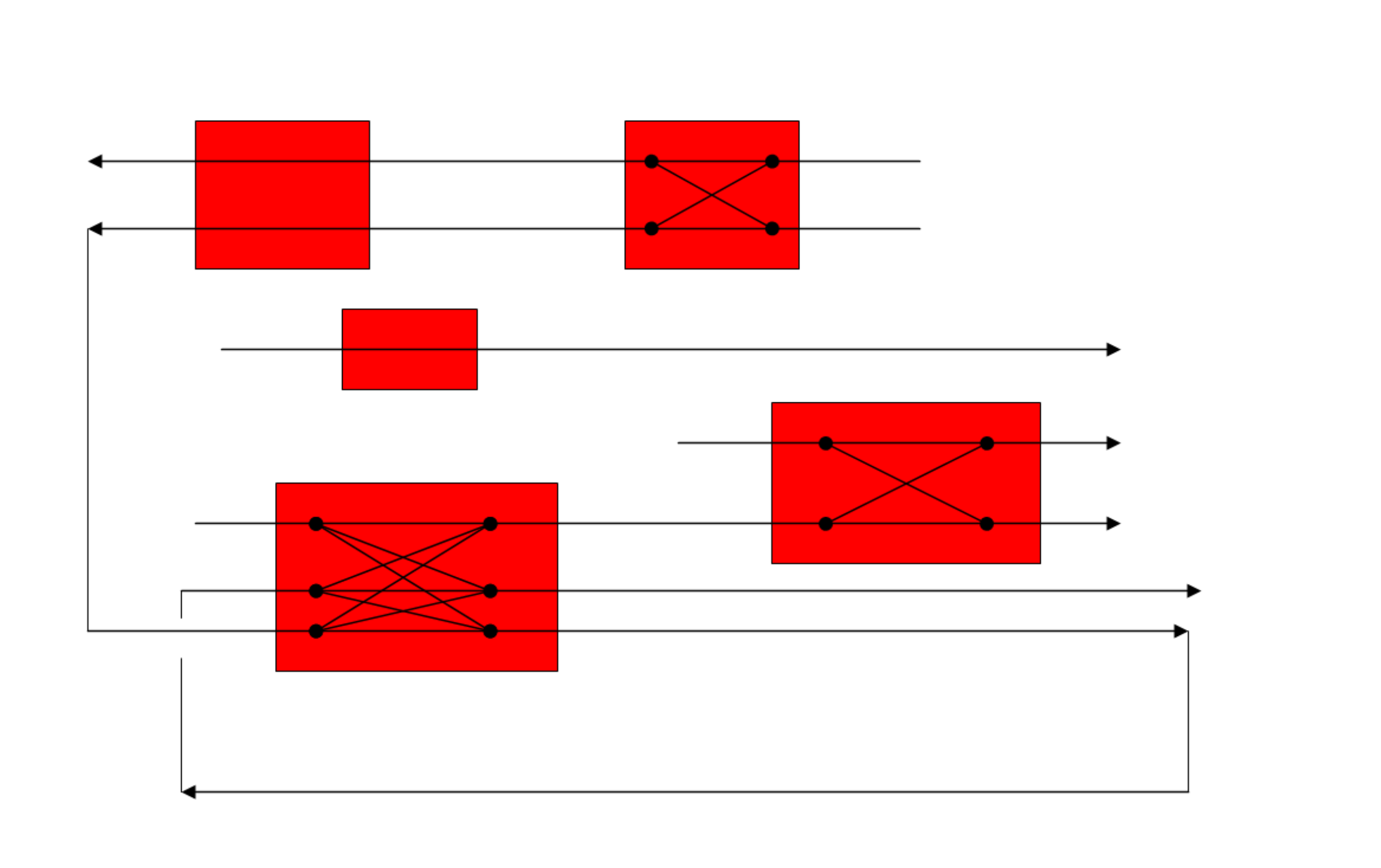}
\caption{(color online) An arbitrary quantum feedback network.}
\label{fig:QTFN_QFN_network}
\end{figure}

If, for instance, we wished to determine the effective SLH model for the
network shown in Fig. \ref{fig:QTFN_QFN_network}, then we would proceed as
follows: first of all we disconnect all the internal lines, this leaves us
with an ``open-loop'' description where all the components are have
independent inputs and outputs, and at this stage we use the parallel sum to
collect all these components into a single open-loop quantum Markov
component; the next step is to make the connections and this involves
feeding selected outputs back in as inputs from the open-loop description,
and to this end we use the feedback reduction formula. This process has
recently been automated using a workflow capture software QHDL \cite{Tezak}
and \cite{Tezak1}. 

\bigskip

\noindent{\textbf{The Two Cavities as an Open Loop}}

\bigskip

The plant and controller form a combined 4-input 4-output system as depicted in Fig. \ref{fig:open_loop} below.
\begin{figure}[h]
	\centering
		\includegraphics[width=0.50\textwidth]{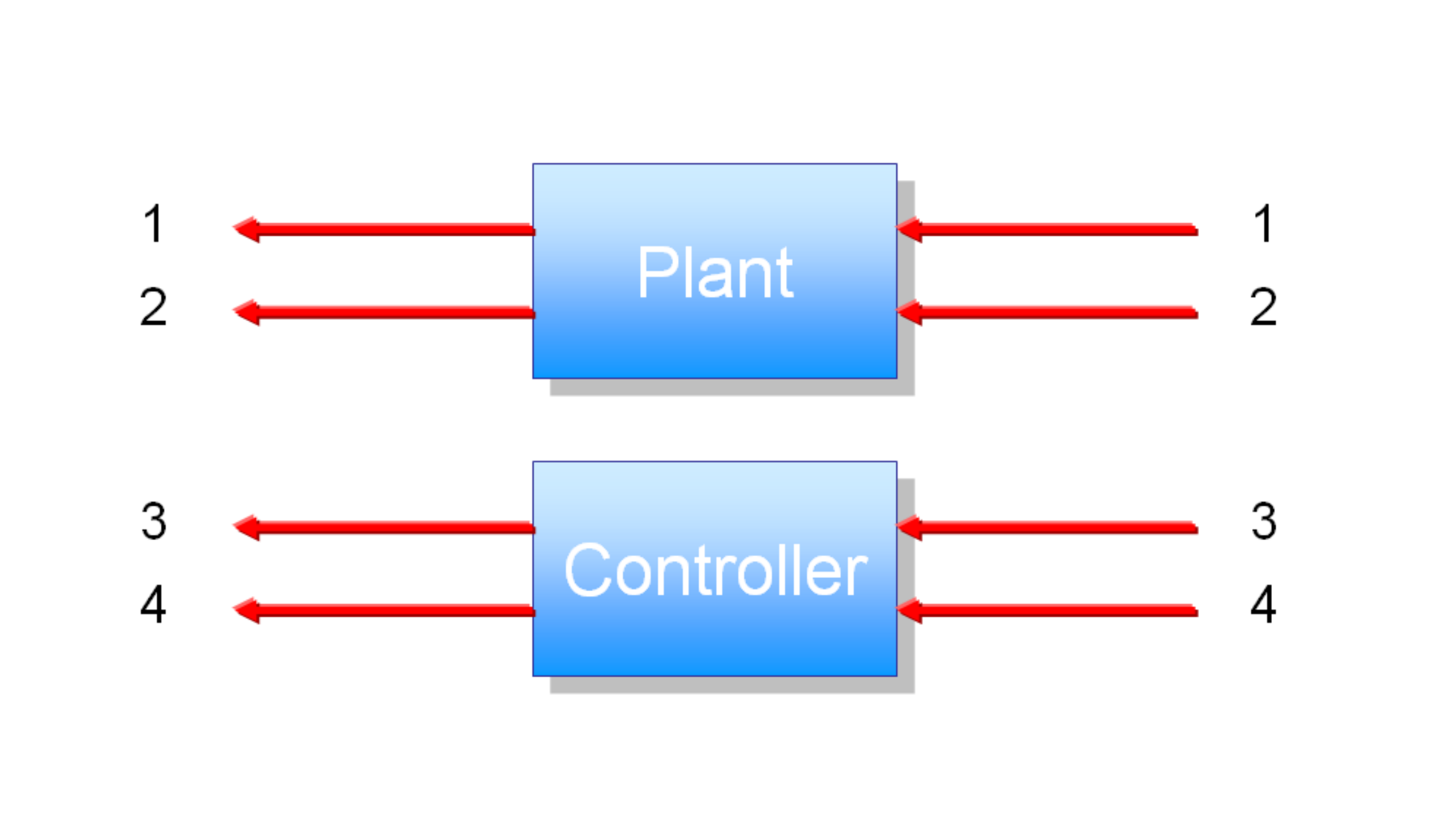}
	\caption{(color online) The plant cavity and controller cavity before the routing connections are made.}
	\label{fig:open_loop}
\end{figure}
Separately they are given by 
\begin{eqnarray*}
\left( S_{p},L_{p},H_{p}\right)  &=&\left( \left[ 
\begin{array}{cc}
1 & 0 \\ 
0 & 1
\end{array}
\right] ,\left[ 
\begin{array}{c}
\sqrt{\frac{\gamma }{2}}a_{p} \\ 
\sqrt{\frac{\gamma }{2}}a_{p}
\end{array}
\right] ,0\right) , 
\end{eqnarray*}
and 
\begin{eqnarray*}
\left( S_{c},L_{c},H_{c}\right)  &=&\left( \left[ 
\begin{array}{cc}
1 & 0 \\ 
0 & 1
\end{array}
\right] ,\left[ 
\begin{array}{c}
\sqrt{\frac{\gamma }{2}}a_{c} \\ 
\sqrt{\frac{\gamma }{2}}a_{c}
\end{array}
\right] ,0\right) .
\end{eqnarray*}

The corresponding open loop system is their parallel sum

\begin{equation*}
\mathsf{V}=\left[ 
\begin{array}{ccccc}
-\frac{\gamma }{2}\left( a_{p}^{*}a_{p}+a_{c}^{*}a_{c}\right) & -%
\sqrt{\frac{\gamma }{2}}a_{p}^{*} & -\sqrt{\frac{\gamma }{2}}a_{p}^{*} & -\sqrt{\frac{\gamma }{2}}a_{c}^{*} & -\sqrt{\frac{\gamma }{2}}%
a_{c}^{*} \\ 
\sqrt{\frac{\gamma }{2}}a_{p} & 1 & 0 & 0 & 0 \\ 
\sqrt{\frac{\gamma }{2}}a_{p} & 0 & 1 & 0 & 0 \\ 
\sqrt{\frac{\gamma }{2}}a_{c} & 0 & 0 & 1 & 0 \\ 
\sqrt{\frac{\gamma }{2}}a_{c} & 0 & 0 & 0 & 1
\end{array}
\right]
\end{equation*}

\bigskip

\noindent{\textbf{Qubit Configuration 1}}

\bigskip

The following pairs $\left( s,r\right) $ consisting of an output port $s$ and
an input port $r$ are to be connected:
\begin{equation*}
\left( 1,4\right),\quad  \left(
4,2\right), \quad \text{and }\quad\left( 2,3\right).
\end{equation*}
The adjacency matrix is then
\[
\eta = 
\left[ 
\begin{array}{ccc}
\eta_{12} & \eta_{13} & \eta_{14} \\ 
\eta_{22} & \eta_{23} & \eta_{24} \\ 
\eta_{42} & \eta_{43} & \eta_{44}
\end{array}
\right]
=\left[ 
\begin{array}{ccc}
0 & 0 & 1 \\ 
0 & 1 & 0 \\ 
1 & 0 & 0
\end{array}
\right]
.
\]
Applying the feedback reduction rules yields
\begin{eqnarray*}
&&\left[ 
\begin{array}{cc}
-\frac{\gamma }{2}\left( a_{p}^{*}a_{p}+a_{c}^{*}a_{c}\right)  & -%
\sqrt{\frac{\gamma }{2}}a_{p}^{*} \\ 
\sqrt{\frac{\gamma }{2}}a_{c} & 0
\end{array}
\right] + \\
&& \left[ 
\begin{array}{ccc}
-\sqrt{\frac{\gamma }{2}}a_{p}^{*} & -\sqrt{\frac{\gamma }{2}}%
a_{c}^{*} & -\sqrt{\frac{\gamma }{2}}a_{c}^{*} \\ 
0 & 1 & 0
\end{array}
\right] \left( \left[ 
\begin{array}{ccc}
0 & 0 & 1 \\ 
0 & 1 & 0 \\ 
1 & 0 & 0
\end{array}
\right] -\left[ 
\begin{array}{ccc}
0 & 0 & 0 \\ 
1 & 0 & 0 \\ 
0 & 0 & 1
\end{array}
\right] \right) ^{-1}\left[ 
\begin{array}{cc}
\sqrt{\frac{\gamma }{2}}a_{p} & 1 \\ 
\sqrt{\frac{\gamma }{2}}a_{p} & 0 \\ 
\sqrt{\frac{\gamma }{2}}a_{c} & 0
\end{array}
\right]  \\
&=& \left[ 
\begin{array}{cc}
-\gamma a_{p}^{*}a_{p}-\gamma a_{c}^{*}a_{c}-\frac{1}{2}\gamma
a_{c}a_{p}^{*}-\frac{3}{2}\gamma a_{c}^{*}a_{p} & -\sqrt{2\gamma }%
(a_{p}+a_{c})^{*} \\ 
\sqrt{2\gamma }(a_{p}+a_{c}) & 1
\end{array}
\right] 
\end{eqnarray*}
from which we see that $L_{1}=\sqrt{2\gamma }(a_{p}+a_{c})$ and the complex
damping is
\begin{eqnarray*}
K_{1}&=&-\gamma a_{p}^{*}a_{p}-\gamma a_{c}^{*}a_{c}-\frac{1}{2}\gamma
a_{c}a_{p}^{*}-\frac{3}{2}\gamma a_{c}^{*}a_{p}\\
&\equiv &
\begin{array}{c}
\left[ a_{p}^{*},a_{c}^{*}\right]  \\ 
\quad 
\end{array}
A_{1}\left[ 
\begin{array}{c}
a_{p} \\ 
a_{c}
\end{array}
\right],
\end{eqnarray*}
with $A_1$ as defined in Section \ref{sec:qubit-write}.

\bigskip

\noindent{\textbf{Qubit Configuration 2}}

\bigskip

The following pairs $\left( s,r\right) $ consisting of an output port $s$ and
an input port $r$ are to be connected:
\begin{equation*}
\left( 1,4\right) , \quad \text{and}\quad \left(
3,2\right) .
\end{equation*}
Applying the feedback reduction rules yields
\begin{eqnarray*}
&&\left[ 
\begin{array}{ccc}
-\frac{\gamma }{2}\left( a_{p}^{*}a_{p}+a_{c}^{*}a_{c}\right)  & -%
\sqrt{\frac{\gamma }{2}}a_{p}^{*} & -\sqrt{\frac{\gamma }{2}}a_{c}^{*} \\ 
\sqrt{\frac{\gamma }{2}}a_{p} & 0 & 0 \\ 
\sqrt{\frac{\gamma }{2}}a_{c} & 0 & 0
\end{array}
\right] \\
&& +  \left[ 
\begin{array}{cc}
-\sqrt{\frac{\gamma }{2}}a_{p}^{*} & -\sqrt{\frac{\gamma }{2}}%
a_{c}^{*} \\ 
1 & 0 \\ 
0 & 1
\end{array}
\right] \left( \left[ 
\begin{array}{cc}
0 & 1 \\ 
1 & 0
\end{array}
\right] -\left[ 
\begin{array}{cc}
0 & 0 \\ 
0 & 0
\end{array}
\right] \right) ^{-1}\left[ 
\begin{array}{ccc}
\sqrt{\frac{\gamma }{2}}a_{p} & 1 & 0 \\ 
\sqrt{\frac{\gamma }{2}}a_{c} & 0 & 1
\end{array}
\right]  \\
&=& \left[ 
\begin{array}{ccc}
-\frac{1}{2}\gamma \left( a_{p}^{*}a_{p}+a_{c}^{*}a_{c}+a_{c}^{*}a_{p}+a_{p}^{*}a_{c}\right)  & -\sqrt{2\gamma }(a_{p}+a_{c})^{*} & -%
\sqrt{2\gamma }(a_{p}+a_{c})^{*} \\ 
\sqrt{2\gamma }(a_{p}+a_{c}) & 0 & 1 \\ 
\sqrt{2\gamma }(a_{p}+a_{c}) & 1 & 0
\end{array}
\right]  
\end{eqnarray*}
from which we see that
\begin{equation*}
S_{2}=\left[ 
\begin{array}{cc}
0 & 1 \\ 
1 & 0
\end{array}
\right] ,
\end{equation*}
$L_{2}=\left[ 
\begin{array}{c}
\sqrt{2\gamma }(a_{p}+a_{c}) \\ 
\sqrt{2\gamma }(a_{p}+a_{c})
\end{array}
\right] $ and the complex damping is
\begin{eqnarray*}
K_{2}&=& -\frac{1}{2}\gamma \left( a_{p}^{*}a_{p}+a_{c}^{*}a_{c}+a_{c}^{*}a_{p}+a_{p}^{*}a_{c}\right) \\
&\equiv &
\begin{array}{c}
\left[ a_{p}^{*},a_{c}^{*}\right]  \\ 
\quad 
\end{array}
A_{2}\left[ 
\begin{array}{c}
a_{p} \\ 
a_{c}
\end{array}
\right],
\end{eqnarray*}
 $A_2$ is as defined in Section \ref{sec:qubit-store}.

\section{Passive open linear quantum input-output systems}
\label{app:passive_linear}
Here we briefly describe the dynamics of passive linear  quantum (input-output) systems using the formalism of quantum stochastic differential equations. In the quantum optical setting, they basically model networks of optical cavities that are connected together by optical fields.  For further details, see, e.g., \cite{GGY08,GZ13}.

Consider a collection of $n$ distinct open quantum harmonic oscillators $a_1,a_2,\ldots,a_n$ satisfying the canonical commutation relations $[a_j,a_k^{*}]=\delta_{jk}$. The oscillators have internal quadratic Hamiltonian of the form $H=\sum_{j,k=1}^{n} (\omega_{jk} a_j^*a_k + c  +{\rm h.c.})$, for some complex constants $\omega_{jk}$ and $c$, and h.c. denotes hermitian conjugate. The oscillators are coupled in a Markovian manner to $m$ external optical fields described by the quantum white noise processes $b_1(t),b_2(t),\ldots,b_m(t)$ from Appendix \ref{app:Configurations}, with all the fields in the vacuum state. The coupling is of the form of the interaction Hamiltonian $H_{\rm int}(t) = \sum_{j=1}^n \sum_{k=1}^m (-\imath K_{jk} a_j^{*} b_k(t) + {\rm h.c.})$ for some complex coupling constants $K_{jk}$. Quanta may also be transferred from  field $k$ to field $j$ via the exchange process $\Lambda_{jk}(t)$, also introduced in Appendix A,  parameterised by a complex unitary $m \times m$ matrix $S$ \cite{HP84,KRP92}. Let $a=(a_1,a_2,\ldots,a_n)^{\top}$, $K_j=[\begin{array}{cccc} K_{j1} & K_{j2} & \ldots & K_{jm} \end{array}]^{\top}$, $S_{jk}=[S_{jk}]_{j,k=1,\ldots,m}$, and $K=[\begin{array}{cccc} K_1 & K_2 & \ldots & K_n \end{array}]$. In the interaction picture with respect to the free field dynamics, and working in units with $\hbar=1$, the joint evolution of the oscillators and the fields is given by a unitary propagator $U(t)$ satisfying the QSDE 
\begin{eqnarray*}
dV(t) &=& \left( -\left(\imath H - \frac{1}{2}a^{*}K^{*}Ka\right)dt +dB(t)^{*}Ka \right. \\
&&\quad \left. \vphantom{\left(\imath H - \frac{1}{2}a^{*}K^{*}Ka\right)dt}  -a^{*} K^{*}S dB(t)^{*} +{\rm Tr}((S-I)^{\top}d\Lambda(t))\right)V(t),
\end{eqnarray*}
with initial condition $V(0)=I$. Here, $B_j(t) = \int_{0}^t b_j(s)ds$, $B(t)=[  B_1(t) , B_2(t) , \cdots , B_m(t) ]^{\top}$, and $\Lambda(t)= \left[\Lambda_{jk}(t)\right]_{j,k=1,2,\ldots,m}$.

The Heisenberg evolution $a(t)=V(t)^* a V(t)$ of the operators in $a$ is given by the QSDE
\begin{eqnarray*}
da(t) &=&  -(\imath \Omega + K^{*}K/2) a(t) dt- K^{*} dB(t),
\end{eqnarray*}
where $\Omega = [\omega_{jk}]_{j,k=1,2,\ldots,n}$, with $\omega_{kj}=\omega_{jk}^{*}$. The input fields $B(t)$ immediately after interaction with the system becomes the output field $Y(t)=U(t)^* B(t) U(t)$ satisfying
\begin{eqnarray*}
dY(t) &=& Ka(t)dt + S dB(t).
\end{eqnarray*}

\section{Controllability and Observability}
\label{app:controllability}

Consider a complex linear (time invariant state-space) system described by the differential equation
\begin{eqnarray*}
\dot{x}(t) &=& A x(t) + Bu(t),\\
y(t)&=& Cx(t) + Du(t),
\end{eqnarray*}
with $A \in \mathbb{C}^{n \times n}$,  $B \in \mathbb{C}^{n \times m}$, $C \in \mathbb{C}^{p \times n}$ and 
$D \in \mathbb{C}^{p \times m}$.

The $\mathbb{C}^n$-valued function $x$ is called the state of the system (state here is in the sense of dynamical systems theory, not in the sense of quantum mechanics), $u$ is a $\mathbb{C}^m$-valued function that is the input to the system, and $y(t)$ is a $\mathbb{C}^p$-valued function that is the output of the system. The dimension $n$ of $x(t)$ is called the state-space dimension of the system. The control system is said to be {\em controllable} if for any time $0<t_f<\infty$ and any state $x(t_f)\in \mathbb{C}^n$ there is a piecewise-continuous $u$ that will bring the system from state $x(0)$ at time $t=0$  to $x(t_f)$ at $t=t_f$. Suppose now that there is no input, $u$ is the zero function, and the initial state $x(0)$ is {\em unknown}. Then the system is said to be observable if for any time $T>0$ the initial state can be determined from observing $y$ over the interval $[0,T]$. Controllability and observability are fundamental properties of linear control systems and an important result in modern control theory is an explicit  necessary and sufficient criterion for verifying these properties. It can be shown that a system is controllable if and only if the controllability matrix $\mathscr{C}$ defined by
$$
\mathscr{C}=[\begin{array}{cccc} B & AB & \ldots & A^{n-1}B \end{array}]
$$
is full rank, and it is observable if and only if the observability matrix $\mathscr{O}$ defined by
$$
\mathscr{O}=[\begin{array}{cccc} C^{*} & A^{*}C^{*} & \ldots & (A^{*})^{n-1}C^{*} \end{array}]^{*}.
$$
is full rank.

The transfer function $G(s)$ of a linear system is a $\mathbb{C}^{p \times m}$-valued complex function defined by:
$$
G(s)=C(sI-A)^{-1}B+D.
$$
Conversely, given a $\mathbb{C}^{p \times m}$-valued complex function $G$, a linear system of some dimension $n$ is said to be a {\em realization} of $G$ if it has $G$ as its transfer function. A linear system that is a realization of $G$ is said to be {\em minimal} if it has the smallest state-space dimension $n$
among all linear systems that have $G$ as their transfer function. Controllability and observability are intimately related to minimality: a realization of $G$ is minimal if and only if the realization is controllable and observable. 

For further details on the contents of this section, see texts on modern control theory such as  \cite{ModernControl}.

\end{document}